\documentclass[twocolumn,showpacs, aps, superscriptaddress, eqsecnum, prd, 
notitlepage, showkeys, nofootinbib, floatfix]{revtex4-2}

\usepackage{graphicx}
\usepackage{dcolumn}
\usepackage{bm}

\usepackage{amssymb}
\usepackage{amsmath}
\usepackage{graphicx}
\usepackage{dcolumn}
\usepackage{hyperref}
\usepackage{color,units}
\usepackage[dvipsnames]{xcolor} 
\usepackage{aas_macros}
\usepackage{lineno}
\usepackage{xspace}
\usepackage[normalem]{ulem} 
\usepackage{float}
\usepackage{orcidlink}
\usepackage{tabularray}

\hypersetup{colorlinks=true, citecolor=teal, urlcolor=teal, linkcolor=teal}

\begin{document}

\title[NSBH and FRBs]{Quantifying the coincidence between gravitational waves and fast radio bursts from neutron star---black hole mergers}
\author{Teagan A. Clarke \orcidlink{0000-0002-6714-5429}}
\email{teagan.clarke@monash.edu}

\affiliation{School of Physics and Astronomy, Monash University, VIC 3800, Australia}
\affiliation{OzGrav: The ARC Centre of Excellence for Gravitational Wave Discovery, Clayton, VIC 3800, Australia}

\author{Nikhil Sarin \orcidlink{0000-0003-2700-1030}}
\affiliation{Oskar Klein Centre for Cosmoparticle Physics, Department of Physics,
Stockholm University, AlbaNova, Stockholm SE-106 91, Sweden} 
\affiliation{Nordita, Stockholm University and KTH Royal Institute of Technology 
Hannes Alfvéns väg 12, SE-106 91 Stockholm, Sweden} 

\author{Eric J. Howell \orcidlink{0000-0001-7891-2817}}
\affiliation{OzGrav, University of Western Australia, Crawley, Western Australia 6009, Australia }

\author{Paul D. Lasky \orcidlink{0000-0003-3763-1386}}
\affiliation{School of Physics and Astronomy, Monash University, VIC 3800, Australia}
\affiliation{OzGrav: The ARC Centre of Excellence for Gravitational Wave Discovery, Clayton, VIC 3800, Australia}

\author{Eric Thrane \orcidlink{0000-0002-4418-3895}}
\affiliation{School of Physics and Astronomy, Monash University, VIC 3800, Australia}
\affiliation{OzGrav: The ARC Centre of Excellence for Gravitational Wave Discovery, Clayton, VIC 3800, Australia}

\begin{abstract}
Fast radio bursts (FRBs) are mysterious astrophysical transients whose origin and mechanism remain unclear. Compact object mergers may be a promising channel to produce some FRBs.
Neutron star---black hole (NSBH) mergers could produce FRBs through mechanisms involving neutron star tidal disruption or magnetospheric disturbances. 
This could present an opportunity for multi-messenger gravitational-wave observations, providing new insight into the nature of FRBs and nuclear matter. 
However, some of the gravitational-wave signals may be marginal detections with signal-to-noise ratios $< 8$ or have large sky location and distance uncertainties, making it less straightforward to confidently associate an FRB with the gravitational-wave signal. 
One must therefore take care to avoid a false positive association.
We demonstrate how to do this with simulated data.
We calculate the posterior odds---a measurement of our relative belief for a common versus unrelated origin of a coincident NSBH and FRB. 
We find that a coincident FRB+NSBH from a common source can yield a statistically significant posterior odds in a network with at least two observatories, but only if we require a coincidence in time and sky location, rather than time alone. 
However, we find that for our model, we require a network signal-to-noise ratio greater than 10 to be confident in the common-source detection, when using a threshold of ln odds $>8$.           
We suggest that a coincident NSBH+FRB detection could help distinguish between FRB engines by discriminating between disrupting and non-disrupting models. 
\end{abstract}

\maketitle

\section{Introduction}
Neutron star---black hole (NSBH) mergers probably make up at least three of the $\gtrsim$ 90 gravitational-wave events observed by the LIGO--Virgo--KAGRA (LVK) collaboration \citep{GWTC2, O3b_population, gwtc3, nsbh_discovery_2021, Gw230529_2024}.\footnote{We refer here to the two unambiguous NSBH detections described in \cite{nsbh_discovery_2021} and the recent NSBH binary merger GW230529 \citep{Gw230529_2024}. However, we note that the primary object of GW230529 has a small probability of being a neutron star rather than a black hole while other ambiguities mean that up to five NSBH events may be included in the latest catalogue of events \citep{gwtc3}.} 
While the first binary neutron star detection was a multi-messenger discovery \citep{BNS_discovery_170817, Abbott2017_multi}, none of the NSBH events observed so far have been confidently associated with an electromagnetic counterpart. 

Fast radio bursts (FRBs) are high-energy, millisecond duration transients in the radio band \citep{Lorimer2007}. 
At least some FRBs are associated with isolated magnetars \citep{CHIME/FRBCollaboration2020}, but there may be multiple channels for producing FRBs \citep[e.g.,][]{Zhang2020}.
Compact object mergers that include neutron stars (binary neutron stars, neutron star---black hole binaries) may  be progenitors of some FRBs \citep[e.g.,][]{Totani2013, Wang2016, Wang2020}, although the FRB rate, at low redshifts, is too high to be accounted for by compact object mergers alone \citep[e.g.,][]{Callister2016, Ravi2019, Luo2020, Chime_o3a_2023}. This channel also does not explain repeating FRBs, further limiting the fraction of FRBs that could come from compact binary mergers.
In order to produce FRBs from a merger remnant, one
needs either a neutron star or black hole central engine. 
A neutron star engine could be either a supramassive neutron star that
collapses some time after the merger or a stable one that never collapses \citep[e.g.,][]{Margalit2019, Margalit_2020, Wang2020}. 
A long-lived neutron star can dissipate energy in different ways, either through magnetic flaring activities or via the blitzar mechanism upon collapse \citep{falcke14, Zhang2014, Ravi2014}. 
For all these dissipation mechanisms, FRB radio emission can be emitted either through coherent emission within or close to the neutron star magnetosphere \citep[e.g.,][]{Katz2016, Kumar2017}, or through emission processes in relativistic shocks that take place far from the magnetosphere, such as synchrotron maser emission \citep[e.g.,][]{Metzger_2017, Lyubarsky_2014}.
We refer readers to e.g., \cite{Zhang2023} for a comprehensive review of the possible FRB progenitors and emission mechanisms.

There have been previous claims of FRB + compact binary merger coincident detections.
Ref.~\cite{Moroianu2023} reported a potential coincidence between the binary neutron star candidate GW190425 \citep{Abbott2020_190425} and FRB 20190425A. However, there are significant observational and theoretical issues with this particular association. First, the inclination angle of the gravitational-wave signal was not consistent with the inclination angle required of the system to produce an FRB that was able to penetrate the optically thick ejecta in the required timescale \citep{Bhardwaj2023, Radice2024}. Second, the lack of an optical kilonova counterpart also disfavours the association \citep{Smartt2024}. Finally, the FRB was launched 2.5 hours after the gravitational-wave signal, which would require the neutron star to have an exotic equation of state to remain stable for this long \citep[e.g.,][]{Abbott2020_190425, MaganaHernandez2024}.
Ref.~\cite{Rowlinson2024} reports the measurement of a coherent radio flash in coincidence with GRB201006A. Although the association for this coincidence is somewhat disfavoured \citep{Sarin2024}, the radio flash, may be evidence of a blitzar \citep{Rowlinson2024, Tian2024}, or a synchrotron maser shock following a binary neutron star merger \citep{Sarin2024}. 
These tentative associations point to the need to have a reliable framework for studying the significance of future associations. 

\begin{figure*}
    \centering
    \includegraphics{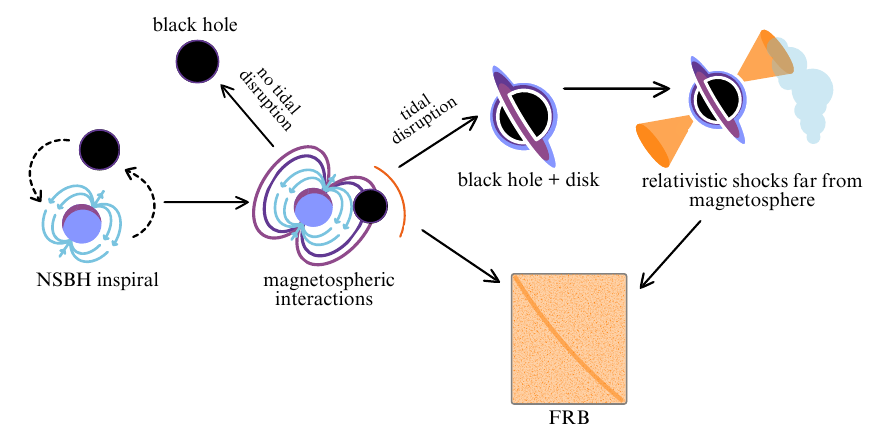}
    \caption{Schematic showing possible pathways to produce an FRB from an NSBH merger. 
    One option involves magnetospheric interactions between the neutron star and black hole. This could include e.g., black hole batteries, charged CBCs or neutron star crust shattering events. For these scenarios tidal disruption of the neutron star is not necessary for producing an FRB.
    These mechanisms are described in detail in Section~\ref{sec:BH_battery}.
    If the neutron star is tidally disrupted, then the system may be able to launch an FRB through mechanisms such as a synchrotron maser shock that occurs far away from the neutron star magnetosphere. 
    These mechanisms are described in detail in Section~\ref{sec:disruption}.}
    \label{fig:schematic}
\end{figure*}

Current low frequency radio facilities provide opportunities to followup on short gamma-ray bursts and gravitational-wave triggers to search for prompt radio transients \citep[e.g.,][]{Yancey2015, Rowlinson2019}.
However, studies attempting to find a link between FRBs and short gamma-ray bursts, implying a compact object merger progenitor, have so far yielded inconclusive results \citep[e.g.,][]{Tian2022, Ashkar2023, Curtin2023, Patricelli2024}. 
If some FRBs are indeed caused by compact object mergers involving neutron stars, then NSBH systems could provide a unique laboratory to probe the engines of the FRB emission. 
The neutron star in an NSBH may or may not disrupt, depending on the binary parameters. 
Both disrupting and non-disrupting NSBH could generate electromagnetic emission \citep[e.g.,][]{Pannarale_2014, Most2023b}, including FRBs.
Figure~\ref{fig:schematic} shows possible configurations to produce an FRB from both a disrupting and non-disrupting system. 

In this paper, we consider simulations of coincident NSBH and FRB triggers. We explore the common source odds statistic for such a coincidence and carefully estimate the component rate and overlap arguments. This is intended to serve as a demonstration of the steps that should be undertaken to confidently associate coincident signals.
This paper is organised as follows: in the next section we discuss the possible mechanisms that could generate an FRB from an NSBH merger. In Section~\ref{sec:odds} we discuss the posterior odds statistic and demonstrate the calculation one would perform for a real coincidence. 
In Section~\ref{sec:simulation} we describe the results of a simulation to illustrate how one may confidently associate a coincident NSBH+FRB and the resulting posterior odds we obtain for these simulations.
In Section~\ref{sec:astrophysics} we explore the astrophysical implications of such a coincidence before discussing the broader implications of our results in Section~\ref{sec:discussion}.

\section{how to make an FRB from an NSBH} 
\label{sec:theory}

\subsection{FRBs from tidally disrupting NSBH mergers}
\label{sec:disruption}
If the black hole in an NSBH is sufficiently light and spinning rapidly, the neutron star may tidally disrupt (see, e.g., \cite{Foucart2020}). Such a disruption could provide a progenitor for a gamma ray burst \citep[e.g.,][]{Mochkovitch_1993,Janka_1999,NAKAR_2007} or kilonova \citep[e.g.,][]{Li_1998, Metzger_2010, kawaguchi_2016}. 

A relatively small fraction of NSBH mergers are likely to disrupt. A low or near-unity mass ratio $q = m_2/m_1 \gtrsim 1/6$, a high prograde black-hole spin ($\chi_\text{BH} \gtrsim 0.5$), or a large neutron-star radius ($R_\text{NS} \gtrsim \unit[12]{km}$) all increase the chances of disruption \citep{Kyutoku_2011, Foucart_2013, Hannam_2013}. These conditions may be relatively rare for binaries that form in isolation---perhaps $\lesssim10\%$ of merging NSBH undergo tidal disruption \citep{Kumar_2017, Zappa_2019, Zhu_2021, Fragione_2021, Sarin2022, Biscoveanu_2023}---although the inferred parameters of the recent NSBH discovery GW230529 \citep{Gw230529_2024} were partially consistent with a disrupting system. 
Additionally, the candidate pulsar+black hole system PSR J0514-4002E \citep{Barr2024} contains an object that is likely a lower mass-gap black hole, although this binary will not merge within a Hubble time. Nonetheless, these are interesting hints that there are disrupting NSBH binaries in nature. 

The disruption of a neutron star in an NSBH merger results in the creation of an accretion disk around the remnant black hole. If the disk is sufficiently massive (a few percent of a solar mass \cite[e.g.,][]{Lee2007, foucart_2012}), highly magnetised, and hot \citep[e.g.,][]{Beckwith2008}, a relativistic jet may be launched. 

The black hole may be surrounded by optically thick dynamical ejecta following the tidal disruption of the neutron star. This may prevent FRB emission from escaping until very late times (tens to hundreds of days after merger), even in a scenario where
the jet clears out a majority of the mass along
the polar funnel \citep{Zhang2014}. 
We propose the most plausible mechanisms for FRB emission following tidal disruption are those where the FRB emission is produced far from the merger remnant via a synchrotron maser \citep{Metzger_2017} or magnetic reconnection \citep{Lyubarsky2020}.
The synchrotron maser scenario requires the accreting black hole central engine to launch a flare, which propagates out
and creates a shock against the upstream jet material, producing radio emission through synchrotron maser instability \citep{Metzger_2019, Margalit_2020}. 
Alternatively, the flare could trigger magnetic
reconnection within the magnetic field inside the
jet producing plasmoids that merge and generate
coherent radio emission, similar to the proposed FRB generation mechanism from luminous accreting X-ray binaries \citep{Sridhar2021}. 

If there is no---or very little---optically thick ejecta or the FRB is launched in the same direction as the relativistic jet then it is plausible that an FRB produced close to the black hole could escape. One way that radio emission could be produced nearby to the black hole is synchro-Compton emission by the high-energy electrons accelerated due to the interaction between the magnetised disk wind and the ambient
medium \citep[e.g.,][]{Usov2000}.

\subsection{Other mechanisms to produce an FRB from an NSBH}
\label{sec:BH_battery}
Interactions related to the neutron star magnetosphere could generate FRB emission regardless of if the neutron star is subsequently disrupted. Electromagnetic flares generated by twisting magnetic flux tubes within the common neutron star+black hole magnetosphere \citep{Carrasco2021} could result in the emission of coherent radio waves \citep{Most2020, Most2023}. The flares depend on the neutron star magnetic field orientation and magnitude, but not the black hole spin \citep{Most2023b}. It is possible that multiple electromagnetic flares may be produced during the binary inspiral, which would get stronger as the binary approaches merger. Such flares have been observed in simulations of binary neutron stars \citep{Most2023}.
Shortly after the merger, strong radio emission could be launched as the magnetic field of the neutron star is expelled away from the black hole \citep[e.g.,][]{Nathanail2020, East2021} in an analogous mechanism to blitzar emission following binary neutron star mergers \citep{falcke14, Most2018}. 

Close to the merger, radio emission could be launched via Poynting flux outflows related to magnetic field line snapping around the black hole \citep[e.g.,][]{D'Orazio2013, Mingarelli_2015, D'Orazio2016}. Ref.~\cite{Mingarelli_2015} proposes the ``black-hole battery'' model for producing an FRB from the merger of a black hole and neutron star. This is an extension of the model proposed by Ref.~\cite{McWilliams_2011} for extracting electromagnetic emission from NSBH to include coherent radio emission. 
According to this model, as the neutron star approaches the black hole, the magnetic field lines of the neutron star thread the event horizon, resulting in the snapping of magnetic field lines, which generate coherent radio emission through magnetic shocks. 
Accumulating charge on the black hole during inspiral could also result in a burst of electromagnetic emission, including in the radio, prior to merger \citep{Dai2019}
Alternatively, a charged spinning neutron star could generate a Poynting flux that generates a relatively weak FRB signal. The signal would be strongest for a highly magnetised, highly spinning neutron star that plunges into a heavy black hole \citep{Zhang2016, Zhang2019}.

Neutron star crust cracking events may be capable of producing FRB emission.
According to this model, the neutron star crust may shatter during the inspiral of an NSBH merger via tidal resonant excitation in the neutron star crust. The crust cracking liberates up to $\sim 10^{47}$ ergs of energy in the final seconds of the NSBH inspiral \citep{Tsang_2012, Tsang2013}. 

We speculate that the large energy budget associated with a crust cracking event could power an FRB if the energy is converted to coherent radio waves, possibly through coupling with the neutron star magnetosphere and the subsequent launch and decay of Alfv\'{e}n waves \citep[e.g.,][]{Kumar2020, Sridhar2021b}. 

\section{quantifying the odds of a common origin}
\label{sec:odds}

We will quantify our relative belief in a common source vs random coincidence of an NSBH+FRB detection by calculating the Bayesian odds. We follow the method described in Ref.~\cite{Ashton2018}. The posterior odds is a ratio of probabilities for two different hypotheses.
Our first hypothesis, which appears in the numerator of the odds, is that the NSBH and the FRB share a common origin.
Our second hypothesis, which appears in the denominator of the odds, is that the NSBH and FRB are independent, and therefore unrelated astronomical events.
The posterior odds can be derived from the ratio between these competing hypotheses and is given by  
\begin{equation}
    \mathcal{O}_{C/R} = \frac{R_\text{NSBH, FRB}}{R_\text{FRB}} \frac{1}{R_\text{NSBH}} \frac{1}{\Delta T} \mathcal{I}_{\Omega}\mathcal{I}_{D_L},
    \label{eq:odds}
\end{equation}
where $R_\text{NSBH, FRB}$ is the rate of NSBH mergers that generate FRBs, $R_\text{FRB}$ is the rate of FRBs and $R_\text{NSBH}$ is the rate of NSBH mergers. 
We will qualitatively explain this equation before providing the mathematical expressions for the various terms in the next subsections.

In Eq.~\ref{eq:odds},
$\mathcal{I}_{\Omega}$ and $\mathcal{I}_{D_L}$ are ``overlap'' integrals \citep{Ashton2018}.
The $\mathcal{I}_\Omega$ overlap describes the degree to which the FRB localisation region overlaps with the gravitational-wave region (see Section~\ref{sec:overlaps} and Eq.~\ref{eq:sky_overlap}).
The $\mathcal{I}_{D_L}$ overlap describes the degree to which the FRB distance posterior overlaps with the gravitational-wave distance posterior (see Section~\ref{sec:overlaps} and Eq.~\ref{eq:dist_overlap}). 
Finally, $\Delta T = \unit[12]{s}$ is the duration of the time window which the LVK searches for coincident FRBs. 
This $\unit[12]{s}$ value is a conventional choice for existing gravitational-wave + FRB coincidence search pipelines \citep[e.g.,][]{Chime_o3a_2023}. The variable $\Delta T$ is chosen somewhat arbitrarily as different models predict different delay times between gravitational-wave emission and FRB emission. Most proposed models, where the FRB is produced close to the merger, predict a relatively short separation time between gravitational-wave trigger and FRB detection $\mathcal{O} ( \unit[1]{sec}$). However, some models, like the synchrotron maser mechanism could support very large delay times, up to a year, depending on the amount of time between the launch of the jet and subsequent shock with out-flowing material \citep[e.g.,][]{Metzger_2017}. We also note that pre-cursor emission models may predict FRB emission at any point during the NSBH inspiral, which lasts for up to $\approx \unit[1]{\text{min}}$ in the LVK observing band \citep[e.g.,][]{Most2020, Most2023, Most2023b}. In the case of a one-year delay, the odds are decreased to $\ll 1$. Although one could opt for a larger time prior, we choose to use the established LVK search-time window since this is the most likely avenue through which potential triggers will be identified. 
We explain the rest of the terms in Eq.~\ref{eq:odds} in the next two sections before discussing our results for $\mathcal{O_{C/R}}$ in Section~\ref{sec:odds}.

\subsection{Overlap integrals}\label{sec:overlaps}
The skymap overlap $\mathcal{I}_\Omega$ is the integral of the gravitational-wave posterior on the sky location multiplied by the electromagnetic spatial posterior divided by the prior and is given by Refs.~\citep[e.g.,][]{Ashton2018, Stachie2020}:
\begin{equation}
    \mathcal{I}_\Omega = \int \frac{p(\Omega | d_\text{GW}) p(\Omega|d_\text{FRB})}{\pi(\Omega)} d\Omega,
\label{eq:sky_overlap}
\end{equation}
where $d_\text{GW}$ and $d_\text{FRB}$ are the sky localisation data for the gravitational-waves and FRB respectively and $\Omega$ is the sky location of the source.
We consider a uniform all-sky prior and a prior scaled by the FRB radio detector's field of view: $\pi(\Omega) = 1/(4\pi f_\text{FoV})$ \citep{Pillas2023}, where $f_\text{FoV}$ is the fraction of the sky spanned by the radio telescope's field-of-view. Simplifying $\pi(\Omega)$, the scaling factor simply becomes the field-of-view in steradians of the telescope that detects the FRB. 

The distance overlap $\mathcal{I}_{D_L}$ is given by
\begin{equation}
    \mathcal{I}_{D_L} = \int \frac{p(D_L | d_\text{GW}) p(D_L|d_\text{FRB})}{\pi(D_L)} d D_L,
\label{eq:dist_overlap}
\end{equation}
where $D_L$ is the luminosity distance to the source.

While the luminosity distance posterior  for the gravitational waves ($p(D_L | d_\text{GW})$) comes directly from the gravitational-wave parameter estimation, the FRB luminosity distance posterior ($p(D_L|d_\text{FRB}$)), is less straightforward to calculate, since the information encoded in the FRB relates to the dispersion measure, rather than the luminosity distance directly. 
We simulate luminosity distance data for the FRB by relating the luminosity distance to the cosmic dispersion measure $\text{DM}_\text{cosmic}$. 
The prior on $\text{DM}_\text{cosmic}$ is parameterised by $\Delta \equiv \text{DM}_\text{cosmic}/\langle\text{DM}_\text{cosmic}\rangle$.
The denominator is calculated using the Macquart relation \citep{Macquart2020}, which is given by 

\begin{equation}
    p_\text{cosmic}(\Delta) = A\Delta^{-\beta} \ \text{exp}\bigg[ - \frac{(\Delta^{-\alpha}-C_0)^2}{2\alpha^2\sigma^2_\text{DM}}\bigg],
\label{eq:macquart}
\end{equation}
with $\sigma_\text{DM} = 0.2z^{-0.5}$ and $[\alpha,\beta] = 3$ and $\langle\text{DM}_\text{cosmic}\rangle$ is given by Equation~2 in Ref.~\cite{Macquart2020}. Here, $C_0$ is determined by requiring that $\langle \Delta \rangle = 1$. We sample from a distribution of $\Delta$ drawn from Eq.~\ref{eq:macquart} and convert the samples to a distribution in luminosity distance.  
In principle, one should account for dispersion measure contributions from the local galaxy (Milky Way + Halo) and the FRB host galaxy \citep[e.g.,][]{Chime_o3a_2023}. However, the Milky Way and halo contributions to the dispersion measure are known functions of the direction to the FRB, and can therefore be adequately accounted for using models of the Milky Way and halo electron density \citep[e.g.,][]{Cordes2002, Yao2017}, , although the halo contribution may be more uncertain due to modelling errors \citep[e.g.,][]{Cook2023}. We assume that the local contribution imparts minimal error on our inference on the distance. The host galaxy contribution is more difficult to quantify. NSBH systems are thought to receive a kick from the component supernovae, such that the eventual merger may take place offset from the host galaxy \citep[e.g.,][]{Narayan1992}. These offsets may be consistent with offsets seen in the short gamma-ray burst population \citep[e.g.,][]{Berger2005, Fox2005, Bloom2006, Troja2008}, although we note that most short gamma-ray bursts likely come from binary neutron stars, not NSBH systems, which might receive smaller kicks due to black holes imparting smaller natal kicks than neutron stars \citep[e.g.,][]{Fryer2012}. The offset means the NSBH merges in a low density environment, resulting in a small host galaxy dispersion measure contribution. However, if there is kilonova ejecta present in the merger environment from the NSBH undergoing tidal disruption, then the resultant contribution from the ejecta matter could be very large \citep[e.g.,][]{Radice2024}. This issue would be mitigated if the FRB is launched along the same axis as a gamma-ray burst jet which would clear out the environment in the direction of the jet \citep{Zhang2014} and reduce the host dispersion measure along the line of sight.
Since there is no simple way to prescribe the host galaxy and/or ejecta contribution to the dispersion measure, we assume that the contribution is small and focus solely on the contribution from $\text{DM}_\text{cosmic}$, which is well-defined by the Macquart relation. 

\begin{figure}
    \centering
    \includegraphics{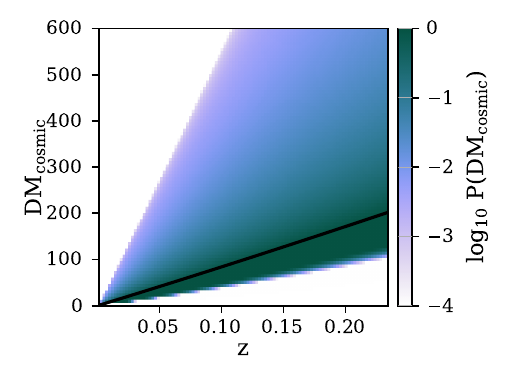}
    \caption{The Macquart relation (Eq.~\ref{eq:macquart}) provides a probabilistic relationship between the cosmic dispersion measure from the intergalactic medium and redshift of a population of FRBs. The black line represents the average redshift for an FRB with a given cosmic dispersion measure. }
    \label{fig:macquart}
\end{figure}

The distance overlap prior $\pi(D_L)$ is calculated using Eq.~\ref{eq:macquart} and is shown in Figure~\ref{fig:macquart}. We sample from this distribution to simulate a luminosity distance posterior for the FRB measurement. We then calculate the overlap between this distribution and the luminosity distance posterior obtained from gravitational waves. We use a common source prior that is uniform in the source frame when calculating the overlap. 

\section{Simulation}
\label{sec:simulation}

We consider three NSBH signals injected into the LIGO Livingston and Hanford observatories at design sensitivity \citep{adv_ligo_2015, AdvancedVirgo, 2020_Kagra} with network SNR $\approx$ 15, 10, and 8.
We choose parameters for our simulations that are similar to the recently discovered binary system PSR J0514-4002E, which contains a pulsar and a compact object within the lower mass gap \citep{Barr2024}. 
A system like this, if it is an NSBH, would provide a plausible pathway for tidal disruption, although we note that non-disrupting systems may also produce FRBs (see Section~\ref{sec:BH_battery}). We choose the neutron star mass of $1.4M_\odot$ and tidal deformability parameters that result in a neutron star radius of $\unit[12]{km}$ that will undergo tidal disruption with a black hole of our chosen parameters.

Table~\ref{tab:inj_params} lists the gravitational-wave parameters chosen for this study.  
We inject and recover the signals using the Bayesian inference library \texttt{Bilby} \citep{Ashton_2019_bilby, Romero_Shaw_bilby} and the \texttt{dynesty} nested sampler \citep{dynesty}. 
We inject the signals into Gaussian noise coloured by the LIGO amplitude spectral density noise curves at design sensitivity.\footnote{We use the \texttt{aligo\_O4high.txt} amplitude spectral density curve taken from https://dcc.ligo.org/LIGO-T2000012/public \citep{ALIGO_curves}} We sample from uniform priors in the chirp mass, mass ratio, component spins, luminosity distance and neutron star tidal deformability.\footnote{We sample with 1000 live points, phase and time marginalisation turned on, and a stopping criterion of $ \Delta \text{log}\mathcal{Z} < 0.1$, where $\mathcal{Z}$ is the Bayesian evidence.}
We use the binary neutron star waveform \texttt{IMRPhenomPv2\_NRTidalv2} \citep{Dietrich2019_tidalv2}, setting the primary component tidal deformability to 0 as expected for a black hole. 
We speed up our inference by employing reduced-order-quadrature in our likelihood evaluations \citep{Canizares2015, Smith2016, Morisaki2023}. 

\begin{table}
    \centering
    \begin{tabular}{c c c}
    \hline
    \hline
        parameter & abbreviation & value \\
        \hline
         black hole mass & $m_1$ & 2.5 M$_\odot$ \\
        neutron star mass & $m_2$ & 1.4 M$_\odot$ \\
         black hole spin & $\chi_1$ & 0.89 \\
         neutron star spin & $\chi_2$ & 0.01 \\
         declination & Dec & 0.36 rad \\
        right ascension &  RA & 1.14 rad \\
         neutron star tidal deformability & $\Lambda_\text{NS}$ & 391 \\
        inclination & $\theta_\text{JN}$ & 0.4 rad \\
        geocent time & t$_c$ &  1369419318.769 s\\

    \hline
    \hline
    \end{tabular}
    \caption{Source-frame parameters for our injected NSBH system. We inject three different signals with aligned spins and identical parameters except for the luminosity distance, which are set to 350, 550 and 750 Mpc to produce signals with network SNRs of 15, 10 and 8 respectively.
    }
    \label{tab:inj_params}
\end{table}

\begin{figure}
    \centering
    \includegraphics[trim={0cm 0cm 0cm 3cm}, clip]{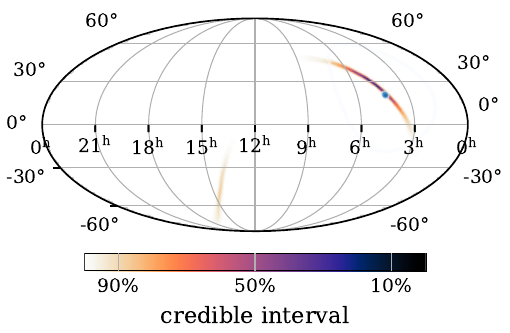}
    \caption{Skymap showing a simulated NSBH system that generates an FRB. The colours represent the gravitational-wave sky localisation posterior. In this simulation we assume a luminosity distance of 350 Mpc, which implies a gravitational-wave network SNR=15. The blue dot represents the $1\sigma$ FRB sky localisation of $\approx \unit[1]{\text{deg}^2}$, assuming the FRB was detected by CHIME. }
    \label{fig:skymap}
\end{figure}

We assume a coincident FRB with a de-dispersed event time matching the merger time of the gravitational-wave signal, i.e., the merger and FRB occur simultaneously in the source frame. We assume the FRB is detected with the Canadian Hydrogen Intensity Mapping Experiment (CHIME) observatory \citep{Newburgh2014SPIE, Bandura2014SPIE, chime_url}. 
Figure~\ref{fig:skymap} shows the gravitational-wave skymap obtained from gravitational-wave inference and the FRB localisation for a FRB detected with CHIME for the SNR=15 simulation. 
We note that the CHIME sensitivity is not uniform across the field of view and that the exposure time depends on declination. For this simulation, the FRB is located in the region of Galactic longitude equating to 10-20 hours exposure time, which is on the order of the baseline sensitivity of CHIME (see Fig.~4 and~5 of \cite{CHIME/FRBCollaboration2021}).

\subsection{Rates and Prior odds}\label{sec:priors}
The prior odds of our calculation represents the probability that an FRB and NSBH come from a common source, before including the overlap statistics of the events. We take the prior odds as the first chunk of Eq.~\ref{eq:odds}, such that
\begin{equation}
    \pi_{C/R} = \frac{R_\text{NSBH, FRB}}{R_\text{FRB}} \frac{1}{R_\text{NSBH}} \frac{1}{\Delta T}.
    \label{eq:prior_odds}
\end{equation}
To calculate $\pi_{C/R}$, we need to estimate the rate of FRBs that have an NSBH progenitor.
With no prior statistically robust detections, we do not know the value of $R_\text{NSBH, FRB}$. 
For this calculation, we assume that we see one coincident event in one year of co-observing time with the LIGO Hanford and Livingston detectors operating at LVK design sensitivity and the CHIME radio observatory. 
In reality, the observing time before obtaining a real coincidence could be shorter or longer, but we choose this rate---assuming a real coincidence---for our demonstration for self consistency.
We calculate the rate of FRBs that are created from merging NSBH assuming one coincidence observed. In future, the circumstances of a real candidate event would provide the rates to use to perform the calculation.  
We note that for this proof of principle study, we are calculating the prior odds using intrinsic rates. We assume that at relatively low redshifts the FRB and gravitational wave rates remain constant. Future studies involving real events should consider using detection rates rather than intrinsic rates.

We take the inferred merger rate of a sub-population of NSBH mergers with similar parameters to GW230529, i.e., where the black hole component resides in the lower mass gap, of $\unit[55] {Gpc^{-3}\,yr^{-1}}$ \citep{Gw230529_2024}. 
This rate is consistent with previous constraints on the NSBH merger rate \citep[e.g.,][]{Mapelli2018, nsbh_discovery_2021}.
We assume that the SNR=8 trigger occurs at the horizon distance for detecting a mass-gap NSBH system at design sensitivity. Because we choose optimal parameters for our simulation, the true volume range will be substantially less than this trigger, as most configurations in location and time would not be detectable at $\unit[750]{Mpc}$.

We use the method described in Ref.~\cite{Chen2021} to estimate the redshifted detectable volume for a system with the source-frame masses listed in Table~\ref{tab:inj_params}, by sampling over the trigger time, sky location and inclination of a binary with our chosen parameters and measuring the network SNR of our samples. 
We choose an SNR threshold of 7 as the limit for triggers to be recorded as sub-threshold triggers, that could still be utilised with an FRB to gain confidence in an association.
The resulting effective co-moving horizon distance for our binary parameters is $\unit[370]{Mpc}$. Assuming the merger rate is uniform in co-moving volume at low redshifts, we find the corresponding rate obtained from this rate and distance is $R_\text{NSBH} = \unit[11] {yr^{-1}}$. Using this effective horizon distance and our assumption of observing 1 FRB+NSBH in a year of co-observing time, we take a rate of FRB+NSBH of $R_\text{NSBH, FRB}=\unit[1]{\text{yr}^{-1}}$ to $\unit[370]{Mpc}$ or $\unit[5]{Gpc^{-3}\text{yr}^{-1}}$.

We assume a rate of FRBs of $\unit[10^4]{Gpc^{-3}\,yr^{-1}}$ \citep{Luo2020}, which corresponds to a rate per unit time of $\unit[1600]{yr^{-1}}$, assuming a uniform in co-moving volume rate to $\unit[370]{Mpc}$. 
Combining these rates with $\Delta T = \unit[12]{s}$ gives a prior odds of $\pi_{C/R} = 144$.

The rate $R_\text{NSBH, FRB}$ is significantly less than the total rate of FRBs ($\approx$ 0.06 \%, of our assumed FRB rate).
Our rate ($R_\text{NSBH, FRB} = \unit[1]{yr^{-1}}$ or $\unit[5] {Gpc^{-3}\,yr^{-1}}$) implies that approximately ten percent of the sub-population of mass-gap black hole NSBH mergers could launch an FRB. 
If we assume that mass-gap black holes make up approximately 60\% of the total merging NSBH population \citep{Gw230529_2024}, then we estimate that up to six percent of all NSBH mergers may launch FRBs. 
This is consistent with some estimates on the fraction of systems that undergo tidal disruption and is e.g., within the 90\% credible interval for the fraction of electromagnetically bright NSBH mergers inferred by e.g., \cite{Biscoveanu_2023} and \cite{Gw230529_2024}. 
We note that some NSBH mergers could be electromagnetically bright without launching an FRB, which would increase this fraction, and that some NSBH systems may launch an FRB without undergoing tidal disruption (see, e.g., Section~\ref{sec:BH_battery}). 

Additional constraints may be placed based on the predicted energies of FRBs from NSBH mergers. Neutron star + black hole events are predicted to produce radio transients with luminosity greater than $\unit[10^{44}]{\text{ erg s}^{-1}}$. The rate of FRBs predicted with luminosities higher than this is $ \unit[\sim 100]{\text{Gpc}^{-3}\text{yr}^{-1}}$ \citep{Luo2020}. At this energy level, we find that NSBH mergers may contribute $ > 5\%$ of the FRB population.
 
\begin{figure}
    \centering
    \includegraphics{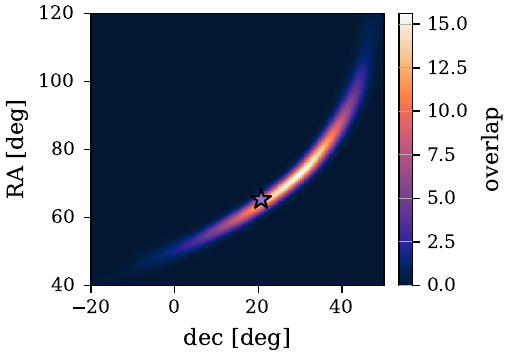}
    \caption{
    Spatial overlap for an SNR=15 NSBH and coincident FRB as a function of RA and Dec. We keep the gravitational-wave posterior fixed and shift the position of the FRB over the grid and calculate the spatial overlap at each point. The ``true'' FRB location is shown as a star. These overlap values have been scaled by the CHIME field-of-view of $\unit[200]{\text{deg}^2}$.}
    \label{fig:sky_overlap_grid}
\end{figure}
\begin{figure}
    \centering
    \includegraphics{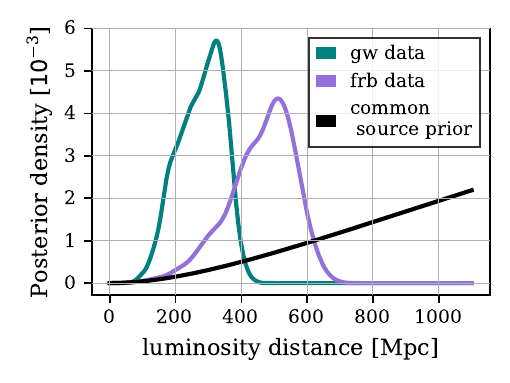}
    \caption{The luminosity distance distributions for the SNR=15 simulation. The teal curve represents the posterior distribution obtained from gravitational-wave parameter estimation. The purple curve shows the simulated posterior for the FRB luminosity distance, where the distribution is obtained by sampling in Eq.~\ref{eq:macquart}. The black curve shows the common-source prior distribution, which we use to calculate the distance overlap. }
    \label{fig:distance_overlap}
\end{figure}

\subsection{Odds Results}
We calculate the odds of a common source origin for each of our simulations, described in Section~\ref{sec:simulation}. We use the prior odds calculated in Section~\ref{sec:priors} and use Eq.~\ref{eq:sky_overlap} and Eq.~\ref{eq:dist_overlap} to calculate the sky and distance overlaps for each of the simulations. 

\begin{figure}
    \centering
    \includegraphics[width=\columnwidth]{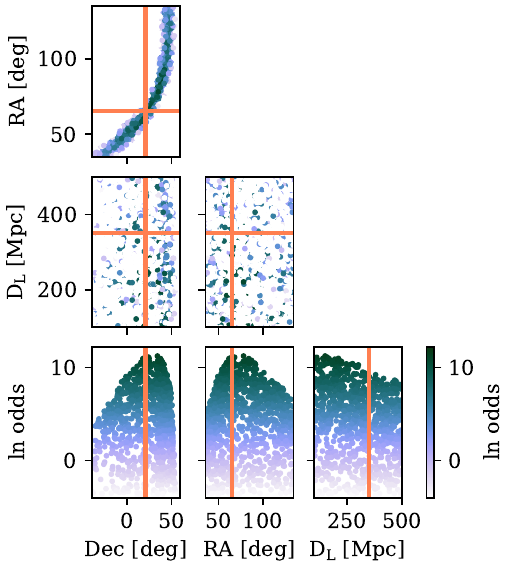}
    \caption{The natural log of the odds for a SNR=15 gravitational-wave injection with a CHIME FRB as a function of overlap parameters: RA, Dec, luminosity distance. The injected sky location and distances are marked with solid lines. We construct this plot by calculating the odds (Eq.~\ref{eq:odds}) for random samples drawn from uniform distributions in RA, Dec and luminosity distance. The plot appears speckled in the luminosity distance parameter space because the panels are 2d slices in a 3d parameter space, and the distance is not correlated with the RA or Dec. Hence the distribution of odds in these panels appears random.}
    \label{fig:odds_corner}
\end{figure}

Figure~\ref{fig:sky_overlap_grid} shows the angular overlap for an FRB and NSBH signal as a function of FRB sky coordinates. The true injected location is highlighted by a star. 
We use the \texttt{ligo.skymap} \citep{LIGO_Skymap} and \texttt{ligo.raven} \citep{Urban2016} python packages to construct the skymaps and numerically calculate the spatial overlap described by Eq.~\ref{eq:sky_overlap}.
For the $\text{SNR} = 15$ simulation we find that $\mathcal{I}_\Omega = 172$ using an all-sky prior and $\mathcal{I}_\Omega = 10.5$ when scaling by the CHIME field-of-view of $\unit[200]{deg^2}$.
We calculate the distance overlap using (Eq.~\ref{eq:dist_overlap}) with the posterior distributions for the luminosity distance obtained using the simulated gravitational-wave data and the simulated FRB distance posterior samples obtained using the procedure described in Section~\ref{sec:overlaps}, using the \texttt{scipy.integrate} python package.
The distributions in luminosity distance for the SNR=15 gravitational-wave signal, simulated FRB and the common source prior are shown in Figure~\ref{fig:distance_overlap}. We find $\mathcal{I}_{D_L}\approx 7$ for our loudest synthetic observation. 
Table~\ref{tab:odds} lists the overlap statistics and the natural log posterior odds $\text{ln } \mathcal{O}_{C/R}$ comparing the common-origin hypothesis to the chance-coincidence hypothesis as per Eq.~\ref{eq:odds}. Figure~\ref{fig:odds_corner} shows the natural log of the odds calculated over a grid of sky localisation and distance parameters.
All calculations assume the same prior odds value, described in Section~\ref{sec:priors}.

We consider $\text{ln }\mathcal{O}_{C/R}=8$ to be strong evidence of a common source event. With this threshold, only the SNR=15 simulation can be confidently identified as a common-source trigger. 
With our model, the odds for the SNR=8 and SNR=10 trigger fall below the characteristic odds considered for a strong association of $\text{ln } \mathcal{O}_{C/R} = 8$. These results suggest that gravitational-wave signals with SNR $>10$ are required to obtain high confidence of an NSBH+FRB significance. However, we caveat that our results are sensitive to the choice of prior probability, particularly the coincident search window time. 

We consider whether a coincidence in time alone is enough to warrant high confidence in a shared origin for two signals. The current CHIME detection rate for non-repeating FRBs is $\approx 1$ per day \citep{CHIME/FRBCollaboration2021}. If there is one FRB and one gravitational-wave signal each day that are independent of each other, occurring in a random sky location each time, then the probability of unrelated signals arriving within 12 seconds of each other is $\approx 10^{-4}$. Over one year of observing, the probability of a coincident in time observation of both signals is $\approx 0.05$---not enough statistical significance for a confident association. Thus, we should not rely on the time coincidence alone to identify coincident events and instead require the additional information of the sky positions, distances and rates to concretely identify a true coincident detection. 
As more FRBs and gravitational waves are detected with improved observatories, the probability of observing signals in the same time window will increase, further highlighting the need for additional information.

We calculate a frequentist $p$-value statistic, which quantifies the probability of a coincident signal randomly arriving at the same time and within the $95\%$ credible interval for sky localisation, such that
\begin{equation}
    p = \left(\frac{\delta t}{\unit[1]{year}}\right)\left(\frac{\delta \Omega }{\unit[4\pi]{sr}}\right),
\label{eq:p_val}
\end{equation}
where the first term is the probability of a chance coincidence in time in one year, and the second term is the chance of an FRB falling within the 95\% credible interval of the gravitational-wave sky localisation area. 
This $p$-value differs from the more standard $p$-value calculation that can be estimated directly from the odds: 
\begin{align}
(1-p)/p = \mathcal{O},
\end{align}
which describes the probability of a single event in isolation being common-source. Our calculation includes the trials factor gained from one year of observations. The $p$-values we calculate are therefore higher than the result one would get using the single event $p$-value statistic. 
Our $p$-value also differs from the one used by Ref.~\cite{Moroianu2023} which uses the data itself to define a coincidence window. This approach can drive $p$-values to arbitrarily low values as one adds more criteria to the calculation. (As a simple example one could look at the probability of e.g., the Sun possessing certain parameters. By including parameters like the temperature, size and age as separate probabilities, the chances of the Sun satisfying all of the criteria becomes very small even if some of the coincidences are not unlikely on their own, e.g., that the Sun is a main sequence star.) Our calculation, in contrast, defines the coincidence criteria independently from the data. This is therefore a more conservative calculation than that of Ref.~\cite{Moroianu2023}.

We summarise these statistics in Table~\ref{tab:odds}. 
Our $p$-values correspond to a detection significance of 3.2, 2.9 and 2.5 $\sigma$ for the SNR=15, 10, and 8 simulated triggers, respectively. These numbers demonstrate moderate confidence in a common origin, but do not reach the gold standard of $5 \sigma$ required for an unambiguous association. 
The odds we calculate for the SNR=15 and SNR=10 simulations are high enough for us to be more confident in the association.
However the odds would not indicate a confident detection for the SNR=8 trigger. We note that the $p$-value and odds can not be directly compared since the odds describes a relative belief between the common source and random chance hypotheses, while the $p$-value quantifies the probability of a random coincidence. 

    \begin{table}
    \centering
    \begin{tabular}{c c c c c}
    \hline
    \hline
        network SNR & $\mathcal{I}_\Omega$ & $\mathcal{I}_{D_L}$ & $\text{ln } \mathcal{O}_{C/R}$ & $p\text{-value}$ \\
        \hline
        15 & 10.5 & 6.6 & 9.2 & $7.1 \times 10^{-4}$ \\
        10 & 6.9 & 2.0 & 7.6 & $1.7 \times 10^{-3}$\\
        8  & 1.1 & 1.1 & 5.1 & $9.1 \times 10^{-3}$\\
    \hline
    \hline
    \end{tabular}
    \caption{ We list, from left to right, the skymap overlap, the distance overlap, the log common origin odds, and the frequentist $p$-value for each of our simulated triggers. Our quoted $p$-values include the trials factors gained from one year of observations. The odds is largely dominated by the on-source time prior of \unit[12]{s}. 
    We take ln odds = 8 as our threshold for a significant detection, only the loudest simulation passes this baseline threshold. This suggests that in some cases the coincidence alone may not be enough to establish a significant common-source origin. }
    \label{tab:odds}
\end{table}

\section{astrophysical implications}
\label{sec:astrophysics}

Observing an FRB with a NSBH merger would provide evidence of a population of FRBs that does not come from magnetars and evidence of multiple channels that produce FRBs.
We now investigate the implications if the FRB resulted from the neutron star tidally disrupting around the black hole.
The tidal disruption would create an accretion disk around the black hole and possibly some relativistic ejecta. 
Since the tidal disruption would not be directly visible in gravitational waves for the SNRs we study here \citep{Clarke_2023}, we can only infer the probability that tidal disruption occurred based on the binary source parameters and knowledge of the neutron star equation of state.
We follow the method used in Refs.~\citep[e.g.,][]{Biscoveanu_2023, Gw230529_2024} to calculate the probability of tidal disruption and the remnant mass if the system underwent tidal disruption. We marginalise over the uncertainty in the neutron star equation of state using the Gaussian process equation of state posteriors obtained by Refs.~\cite{Legred_2021, legred_2022_data} using pulsar and gravitational-wave measurements. We assign each of our gravitational-wave samples to an equation of state posterior sample to calculate the neutron star compactness. 

\begin{figure*}
    \centering
    \includegraphics{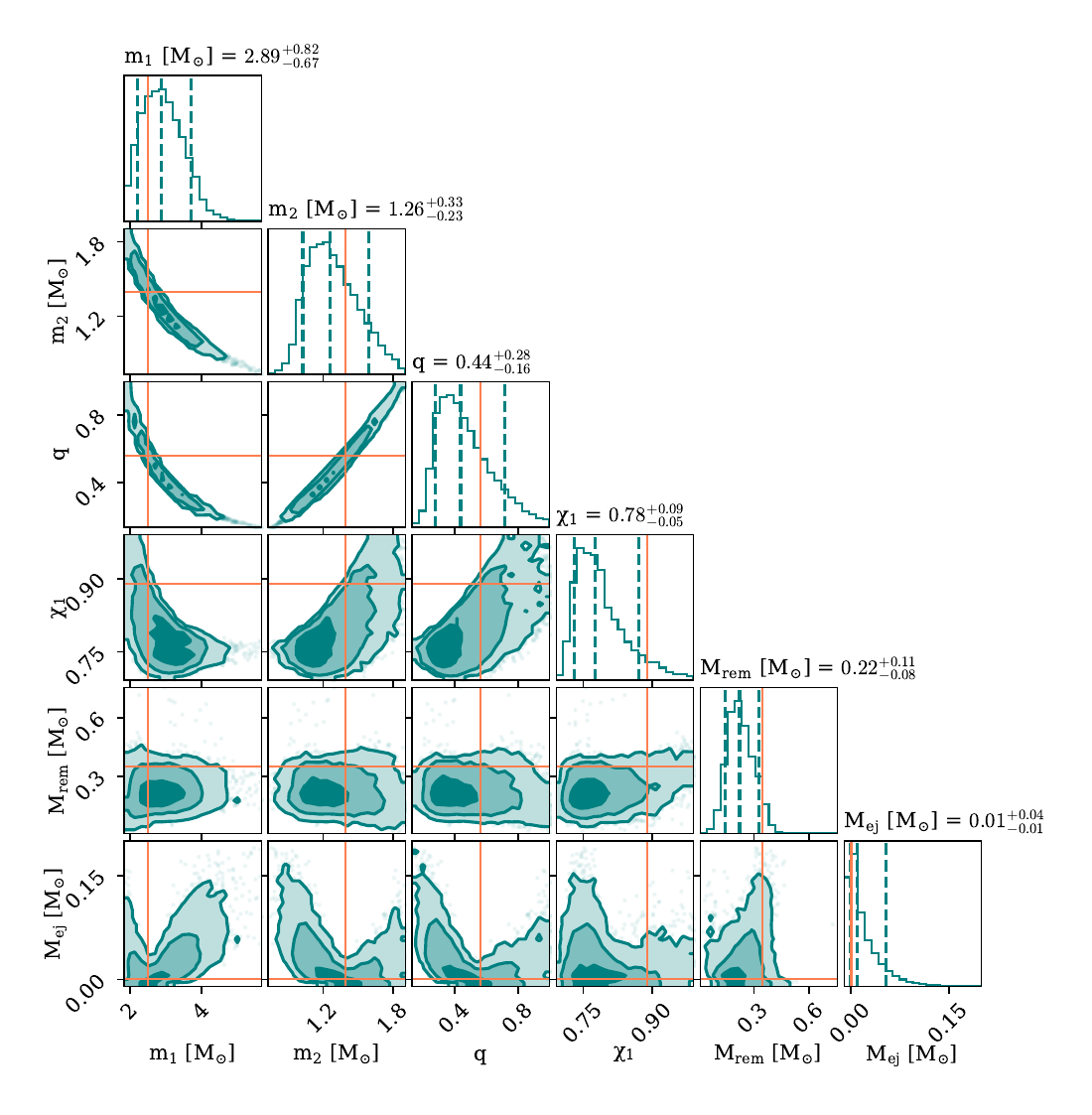}
    \caption{Selected posterior parameters from gravitational-wave parameter estimation of an SNR=15 NSBH injection. The posteriors for the remnant mass and ejecta mass are also shown. The orange lines show the injection parameters. }
    \label{fig:corner_M_rem}
\end{figure*}

We use the fitting formulae, including the spin dependant properties of neutron stars \citep{Cipolletta2015, Breu2016, Foucart2018}, which were used in Ref.~\cite{Biscoveanu_2023} to calculate the remnant baryonic mass remaining outside of the black hole following the merger. The remnant mass is the sum of all of the baryonic material that remains outside of the black hole following the merger. This includes the unbound ejecta and material that remains bound in an accretion disk. The disk mass comprises approximately $80\%$ of the total remnant mass \citep[e.g.,][]{Gottlieb2023}. The remnant mass depends on the mass ratio, equation of state and component spins of the binary \citep{Foucart2018}. We also calculate the ejecta mass, which refers to the material that becomes unbound from the black hole and is launched far from the black hole. We calculate the ejecta mass from the fitting formula derived in Ref.~\cite{kawaguchi_2016}. The inferred parameters from the gravitational-wave analysis and the inferred remnant and ejecta mass obtained by marginalising over the equation of state uncertainty are shown in Figure~\ref{fig:corner_M_rem}.   
The ejected material is important for powering a potential optical kilonova counterpart \citep[e.g.,][]{Li_1998}. 
If the bound disk material is highly magnetised, a relativistic jet can be launched as the material accretes onto the black hole \citep[e.g.,][]{Paschalidis2015}. The timescale for the launched jet after merger and the duration of the jet remain open questions and likely depend on the pre-merger binary parameters. The disk mass likely needs to be relatively high for a jet to be launched ($\gtrsim 0.02 M_\odot$ or at least $5 \%$ of the neutron star mass \citep{Stone2013, Ruiz2018}), although the precise threshold disk mass required to launch a jet is unknown.

Since our simulation is tuned to optimise tidal disruption, we find that all of the samples in our simulation are disrupting, resulting in a probability of tidal disruption of 1. If we follow \cite{Stone2013} and use a threshold disk mass of $0.05 M_\odot$ required to launch a jet, we find that $99 \%$ of our samples satisfy this requirement. 
We also find that $\approx 80\%$ of our samples have a non-zero ejecta mass.
We find an upper limit on the remnant mass of $0.4 M_\odot$ at the $99 \%$ credible interval. 

\section{discussion and conclusions}
\label{sec:discussion}
The physical mechanism behind FRBs remains unknown. While at least some of them seem to come from magnetars \citep{CHIME/FRBCollaboration2020}, there could be multiple channels for producing FRBs \citep[e.g.,][]{Zhang2020}. Compact binary coalescence events, such as NSBH, are one possible FRB sub-channel. 
We explore the scenario where a FRB is generated by an NSBH tidal disruption through a synchrotron-maser emission far from the black hole accretion disk---although a coincident observation from an NSBH whose parameters disfavour tidal disruption could support models of electromagnetic emission that do not require tidal disruption, such as magnetic field re-connection events \citep[e.g.,][]{Most2023}. 

We demonstrate how a coincident NSBH and FRB could be quantified as common-source using the odds statistic. For signals with SNR $>10$, the odds significantly support a common origin. For a weak gravitational-wave signal with SNR $\approx 8$, we find ln odds $\approx 6$, which falls short of achieving the conventional threshold of ln odds = 8. The SNR=10 simulation represents a marginal detection with ln odds = 7.6, suggesting that a confident association requires a gravitational-wave signal with SNR $>10$. However, we note that our results remain prior dependant and the odds could become significant for lower SNR gravitational waves depending on the choices of on-source search time and rates. Our results suggest that in some cases, a true coincidence may not necessarily produce a significant ln odds or frequentist $p$-value. Increased confidence in the gravitational-wave signal may be required, through statistics such as the Bayes coherence ratio \citep[e.g.,][]{Veitch2010}, or a three-detector network being in operation at the time of the event. 

A coincident observation of an FRB with a tidally disrupting NSBH could allow us to place limits on the remnant mass and ejecta mass of the neutron star tidal disruption, by marginalising over the uncertainty in the neutron star equation of state. 

Our results are subject to large uncertainties, particularly in the expected rates of coincident FRB and NSBH, and the expected delay times between the gravitational-wave and FRB signal. With no confirmed detections of FRBs from NSBH, the rates we use in our demonstration are speculative. 
However, we anticipate that the odds statistic will provide a reliable baseline for testing potential associations in the future. This work is intended as a demonstration of the calculations to be performed to establish this baseline for claiming a coincident detection. 
We predict rates of FRB-causing NSBH mergers of up to $5 \ \text{Gpc}^{-3} \text{yr}^{-1}$ if a coincident event is observed during a year-long observing run at design sensitivity. This could correspond to a high fraction of all tidally disrupting NSBH systems, but only $\approx 0.05 \%$ of all FRBs. 
 
A population of binaries that produce FRBs may emerge during the Square Kilometre Array / Cosmic Explorer---Einstein Telescope era. Cosmic Explorer and Einstein Telescope \citep{ Abbott2017_XG, CE, ET} are planned next-generation gravitational-wave observatories that will be able to detect almost every NSBH merger in the observable universe. Meanwhile the future radio observatory called the Square Kilometre Array (SKA) \citep{Huynh2013, Braun2019} is predicted to observe upwards of $\unit[10^6]{ \text{FRBs sky}^{-1} \text{yr}^{-1}}$ \citep{Hashimoto2020}. However, upwards of $10^8$ FRBs occur in the observable Universe every year \citep{Fialkov2017}. Given our assumption that disrupting NSBH systems may account for $\approx 0.05\%$ of FRBs and that SKA will see $\approx 1\%$ of all FRBs in the observable Universe, SKA / Cosmic Explorer may see $\approx 500$ coincident NSBH+FRB per year.

\section{Acknowledgements}
We thank the referees for their helpful suggestions. 
We thank Neil Lu for helpful comments on this manuscript.
We thank Sylvia Biscoveanu for help with equation of state and remnant mass code and Yuri Levin for interesting discussions about this work.
This work is supported through Australian Research Council (ARC) Centres of Excellence CE170100004 and CE230900016, Discovery Projects DP220101610 and DP230103088, and LIEF Project LE210100002. 
T. A. C. receives support from the Australian Government Research Training Program.
N.S. is supported by a Nordita Fellowship and also acknowledges support from the Knut and Alice Wallenberg foundation through the ``Gravity Meets Light'' project. Nordita is supported in part by NordForsk. 
The authors are grateful for for computational resources provided by the LIGO Laboratory computing cluster at California Institute of Technology supported by National Science Foundation Grants PHY-0757058 and PHY-0823459, and the Ngarrgu Tindebeek / OzSTAR Australian national facility at Swinburne University of Technology. 

\section{Data availability}

The data underlying this article will be shared on reasonable request to the corresponding author.

This work made use of the following software packages: \texttt{astropy} \citep{astropy:2013, astropy:2018, astropy:2022}, \texttt{matplotlib} \citep{Hunter:2007}, \texttt{numpy} \citep{numpy}, \texttt{python} \citep{python}, \texttt{scipy} \citep{2020SciPy-NMeth, scipy_11702230}, \texttt{Bilby} \citep{Ashton_2019_bilby, Bilby_2602178}, \texttt{dynesty} \citep{dynesty} and \texttt{corner.py} \citep{corner-Foreman-Mackey-2016, corner.py_4592454}.
Parts of the results in this work make use of the colormaps in the CMasher package \citep{2020JOSS....5.2004V, CMasher_10677366}.
Some of the results in this paper have been derived using \texttt{healpy} and the HEALPix package\footnote{http://healpix.sourceforge.net} \citep{Zonca2019, 2005ApJ...622..759G, healpy_12746571}.
Software citation information aggregated using \texttt{\href{https://www.tomwagg.com/software-citation-station/}{The Software Citation Station}} \citep{software-citation-station-paper, software-citation-station-zenodo}.

\bibliography{bib}
\end{document}